# Electrical power dissipation in carbon nanotubes on single crystal quartz and amorphous SiO$_2$

Cheng-Lin Tsai,[1,2] Albert Liao,[3,4] Eric Pop,[3,4,5] and Moonsub Shim[1,2,a]

[1]*Department of Materials Science and Engineering,* [2]*Frederic Seitz Materials Research Laboratory,* [3]*Department of Electrical and Computer Engineering,* [4]*Micro and Nanotechnology Laboratory, and* [5]*Beckman Institute, University of Illinois, Urbana, Illinois 61801, USA*

**Abstract**

Heat dissipation in electrically biased semiconducting carbon nanotubes (CNTs) on single crystal quartz and amorphous SiO$_2$ is examined with temperature profiles obtained by spatially resolved Raman spectroscopy. Despite the differences in phonon velocities, thermal conductivity and van der Waals interactions with CNTs, on average, heat dissipation into single crystal quartz and amorphous SiO$_2$ is found to be similar. Large temperature gradients and local hot spots often observed underscore the complexity of CNT temperature profiles and may be accountable for the similarities observed.

[a] Electronic mail: mshim@illinois.edu

Exceptional electrical and thermal properties make carbon nanotubes (CNTs) excellent candidates as elements of next-generation electronics including high performance semiconductors, electrical interconnects, heat sinks, and nanoscale heaters.[1-3] With the aggressive down-scaling of device dimensions and increase in circuit density, thermal management becomes increasingly important. In addition to useful characteristics such as diameter/chirality and degree of disorder, Raman-active phonon modes can provide insights into the thermal response of CNTs.[4-6] Because of the importance of optical phonon (OP) scattering at high biases, Raman studies have been valuable in understanding non-equilibrium electron transport in both CNTs and graphene.[5-8] For instance, hot OPs as well as CNT-substrate interactions strongly influence high-field electron transport characteristics (e.g., negative differential conductance being observed only in suspended CNTs rather than those resting on an $SiO_2$ substrate[9]). Hence, examining how different substrates influence thermal response is essential for the design of CNT electronics.

Amorphous $SiO_2$ (a-$SiO_2$) as a thermally-grown oxide on heavily doped Si is by far the most common CNT substrate used to date. However, single-crystal quartz has also become important as it can lead to nearly perfect alignment of CNTs during growth.[10] The two substrates have the same constituent atoms, but substantially different thermal conductivities. Furthermore, spontaneous alignment on single crystal quartz has been attributed to strong van der Waals (vdW) interactions along the growth direction,[11] which should, in principle, result in better thermal coupling. These differences and similarities have motivated this study on comparing heat dissipation processes in electrically biased semiconducting CNTs on these two substrates. Furthermore, while temperature profiles of metallic CNTs supported on substrates or suspended have been previously examined,[12,13] similar spatially resolved studies have not been reported on semiconducting CNTs. Non-uniform electric fields expected and sensitivity to local chemical environment especially to substrate surface charges make it even more important for such studies to be carried out on devices incorporating semiconducting CNTs.

Horizontally aligned CNTs were grown by chemical vapor deposition on ST-cut quartz (Hoffman Materials) using ferritin (Sigma-Aldrich) and $CH_4$ as the catalyst and carbon source, respectively.[14] For measurements on a-$SiO_2$ substrates, aligned CNTs grown on quartz were transferred onto Si substrates with thermal oxide (300 nm).[15] Lithographically patterned metal electrodes (2 nm Ti and 50 nm Pd) were deposited to define 4 μm long CNT channels. Scanning electron microscopy (SEM), atomic force microscopy (AFM), and electrical breakdown[16] were conducted to determine length, diameter, and location of failure as well as to ensure only one CNT spanned the channel of interest. Raman measurements were carried out on Jobin-Yvon Labram HR800 using a 100x air objective with 633 nm laser excitation source, and Ar flowing over the samples. The laser spot size was ~ 1 μm and the power was kept at 1 mW.

Figure 1 shows Raman G-band spectra of a semiconducting CNT on quartz under electrical bias ($V_d$). The downshift of the G-band frequency ($\omega_G$) with increasing $V_d$ indicates increasing temperature ($T$) from Joule heating. Due to the higher intensity, we consider only the longitudinal optical phonon mode here. Estimates of $T$ from changes in $\omega_G$ are often made using the calibrated Raman G-band $T$ coefficient, $\chi_G \sim -0.03$ cm$^{-1}$/K.[14,17,18] However, such calibrations are made under equilibrium conditions. Joule heating leads to a non-equilibrium situation where charge carriers and OPs are at a significantly higher $T$ than the lattice.[5,7,8] Therefore, to estimate the lattice $T$, we utilize $d\omega_G/dT_{RBM} = -0.021$ cm$^{-1}$/K given in Ref. 5 where the $\omega_G$ downshift with Joule heating was reported along with the temperature of the radial breathing mode (which should be at equilibrium with the lattice).

Figure 1(b) shows the change in $\omega_G$ and the corresponding change in $T$ as a function of power per unit length ($P$). If we assume $T$ profile to be uniform along the length ($x$) of a 4 µm long CNT, we expect $T(x) = T_o + P/g$, where $T_o$ is the substrate $T$ and $g$ is the thermal conductance per unit length from the CNT.[19] Here, $g$ includes the interfacial CNT-substrate thermal resistance in series with a spreading heat conduction term into the substrate, however the former typically dominates.[16] From the linear fit shown in Fig.1(b), we obtain $g = 0.18$ Wm$^{-1}$K$^{-1}$, similar to previously reported values.[5,19,20] However, an asymmetric $T$ profile with the highest $T$ near the ground electrode (assuming hole transport with positive $V_d$) is expected in semiconducting CNTs due to the non-uniform electric field along their length.[21] Electrical breakdown measurements have recently demonstrated that this is indeed, on average, the exhibited behavior.[16] These expectations and findings point to the importance of direct measurements of $T$ profiles during Joule heating of semiconducting CNT devices.

An asymmetric profile with $T$ drop as large as ~550 K around the midpoint with the highest $T$ occurring near the ground electrode can be seen for a semiconducting CNT on a-SiO$_2$ substrate in Fig. 2(a). When the polarity of the applied $V_d$ is reversed, the maximum $T$ position shifts accordingly indicating that the observed behavior is not an artifact of asymmetric contact resistance.[22] The SEM image taken after all measurements have been carried out and $V_d$ pushed to device failure shows the location of electrical breakdown to be at the expected position of highest $T$. Figure 2(b) shows that $T$ measurements made at only one location can lead to large apparent variations in $g$: the highest $T$ location yields $g = 0.07$ Wm$^{-1}$K$^{-1}$ versus 0.14 Wm$^{-1}$K$^{-1}$ at 1.5 µm from the left electrode. Similarly large discrepancies are shown in Figs. 2(c) and 2(d) for a semiconducting CNT on quartz with $g = 0.06$ Wm$^{-1}$K$^{-1}$ and 0.18 Wm$^{-1}$K$^{-1}$ near and away from the highest $T$ region, respectively.

While 16 out of 24 semiconducting CNTs examined exhibit $T$ profiles similar to those shown in Fig. 2 with the highest $T$ location closer to the ground electrode (Fig. 3 inset), a significant number of CNTs show what appears to be a random distribution of local hot spots. Examples are shown in Fig. 3 and in the supplementary material. These results further emphasize the importance of obtaining $T$ profiles and the necessity of a better approach to extracting $g$ values in order to compare different substrates. Following Ref. 16, we solve the heat diffusion equation along the CNT using two distinct power dissipation profiles $P(x)$ to capture the two types of behaviors. A quadratic $P(x)$ is used for behavior of the type in Fig. 2. A Gaussian with a constant background is used for cases similar to Fig. 3. Solid lines in Figs. 2 and 3 are the fitted $T$ profiles and Fig. 4 shows the extracted $g$ values. We note that the Gaussian profile captures the experimental results better whenever there is a steep $T$ gradient even if the CNT exhibits expected maximum $T$ near the ground electrode. This behavior may indicate that significant number of otherwise "well-behaved" semiconducting CNTs may nevertheless fall within the category with random local hot spots. In fact, the majority of CNTs examined (13 out of 24) are better described as having local hot spots.

To consider how the two different substrates should affect heat dissipation, we compare our results to the diffuse mismatch model (DMM)[23] in Fig. 4. We calculate $g(T)$ using DMM following Ref. 16 for a 1.5 nm diameter CNT (all CNTs examined here have diameter between 1 and 2 nm as measured by AFM). The calculated values include heat spreading into the substrates. While DMM provides only upper limits, quartz is predicted to exhibit better thermal coupling than a-$SiO_2$ as shown in Fig. 4. This may be expected since quartz has higher phonon velocity, thermal conductivity and atomic density. Given that CNTs should have stronger vdW interactions with quartz (along the alignment direction)[11] than with a-$SiO_2$, an even larger difference in interfacial thermal coupling could be expected based on recent molecular dynamics simulations.[24] Surprisingly, we observe no difference between quartz and a-SiO$_2$ substrates in Fig. 4 within experimental error margins (average $g$ = 0.10 ± 0.02 Wm$^{-1}$K$^{-1}$ is obtained with each substrate for this range of CNT diameters).

There are several mechanisms which may be consistent with the similar thermal coupling observed. Surface phonon polariton (SPP) scattering was theoretically suggested to be an important energy dissipation pathway.[25] Quartz and a-SiO$_2$ should have similar SPPs, which would then lead to similar thermal coupling. In fact, calculations based on quartz have been used to describe CNT devices on a-SiO$_2$ substrates.[26] However, the SPP mechanism should be sensitive to the vdW distance, and stronger vdW interactions expected on quartz[11] should lead to better thermal coupling. Furthermore, local hot spots we observe cannot be explained by the SPP mechanism. Even the more common cases, where the maximum $T$ occurs near the ground electrode, are often better described by Gaussian profiles suggesting that these may also be random local hot spots that happen to be near the ground electrode.[22]

Adsorbed molecules such as hydrocarbons, water and oxygen from the ambient may provide alternate and parallel heat dissipation pathways.[27] Given the same ambient gas environment, possible molecular pathways should be similar for the two substrates and such a mechanism would be consistent with local hot spots – i.e. random distributions of adsorbed molecules. However, at least for hydrocarbons and surfactants, thermal coupling between CNTs and "soft" molecules has been shown to be quite poor[28,29] and, based on our measured values of $g$, direct coupling to the substrate should be about an order of magnitude better.

Charge trapping in the substrate surface may also play an important role. Substrate charging has been shown to cause large hysteresis in CNT transistors[30,31] and such charges can lead to large local electric fields that may be responsible for the observed local hot spots. Localized substrate charging may even introduce electrostatic forces that can strongly alter local interactions (e.g., a trapped hole would repel net positively charged p-type CNT). Such a scenario is consistent with random locations undergoing high degree of local heating leading to similar behavior on the two substrates.

Finally, we comment on the possible role of defects on the heat dissipation process. All CNTs examined here exhibit very little or no observable D-band in their Raman spectra. Intentional introduction of defects by covalent sidewall functionalization with 4-bromobenzene diazonium tetrafluoroborate does not significantly alter the heating behavior. This result can be explained by the "healing" effect of Joule heating as evidenced by the loss of the D-band upon electrical biasing.[22]

In summary, we have shown that single crystal quartz and a-SiO$_2$ substrates exhibit similar thermal coupling to semiconducting CNTs. A significant number of CNTs exhibit unexpected local $T$ spikes that may be explained by substrate surface charges leading to large local electric fields and possibly altering the local interactions. Understanding such unexpected heating behavior is especially important in devising efficient thermal management schemes for nanoscale devices.

This material is based upon work supported in part by the MSD Focus Center, under the Focus Center Research Program (FCRP), a Semiconductor Research Corporation entity and in part by the NSF grants 09-05175 and CAREER 09-54423. Experiments were carried out in part in the Frederick Seitz Materials Research Laboratory Central Facilities, University of Illinois.

**Figure Captions**

FIG. 1. (Color online) (a) G-band Raman spectra of a 4 μm long semiconducting CNT on quartz at the indicated source-drain bias ($V_d$). (b) Change in G-band frequency ($\Delta\omega_G$) and the corresponding temperature as a function of power per unit length. Solid line is a linear fit.

FIG. 2. (Color online) Temperature profiles from measured G-band frequency shift ($\Delta\omega_G$) at the indicated electrical biases and corresponding SEM images taken after electrical breakdown for a CNT on quartz (a) and on a-SiO$_2$ (c) substrates. Solid curves for the highest bias cases are the fitted temperature profiles from solving the heat diffusion equation (Ref. 16) using quadratic power profiles. SEM images are scaled same as the position axes. Ground electrode is on right for both CNTs. Power dependence of G-band frequency shift and the corresponding temperature change at the locations indicated by the arrows in the SEM images for the same CNTs on quartz (b) and a-SiO$_2$ (d). Filled squares (circles) correspond to black (blue) arrow locations. Solid lines are linear fits.

FIG. 3. (Color online) Temperature profiles at the indicated electrical bias (ground electrode on the right side) for a CNT showing an unexpected heating behavior. Solid curve is the fitted profile from solving the heat diffusion (Ref. 16) equation using a Gaussian power profile. Histogram of the highest *T* locations for all CNTs examined is shown in the inset.

FIG. 4. (Color online) Comparison of experimental (symbols) and calculated (lines) values of interfacial thermal conductance including heat spreading into the substrates (*g*) on quartz (black) and a-SiO$_2$ (red). $T_{max}$ is the maximum temperature experimentally measured. Error bars reflect ± 0.005 cm$^{-1}$K$^{-1}$ uncertainty in G-band *T* coefficient often reported (see Ref. 14).

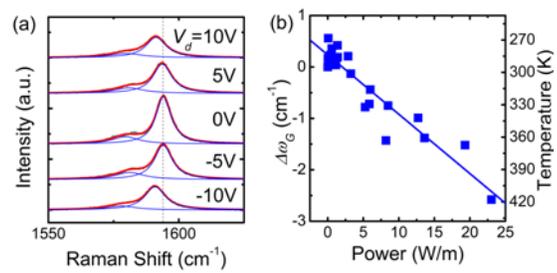

Figure 1

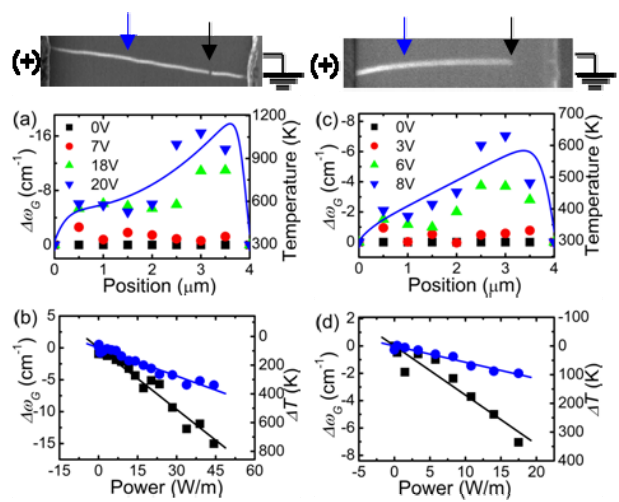

Figure 2

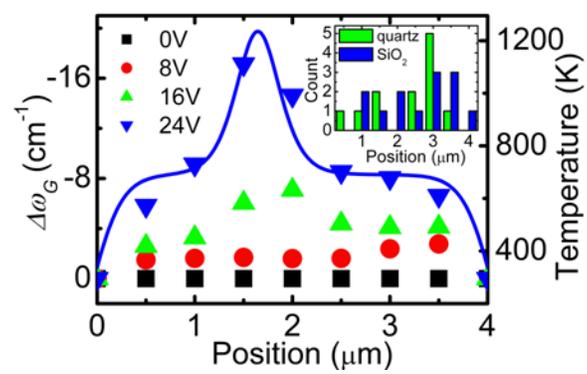

Figure 3

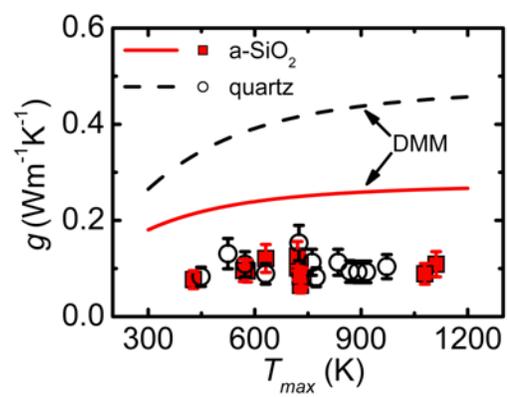

Figure 4